\documentclass{ws-ijmpd}

\begin{document}

\markboth{Janilo Santos, Marcelo J. Rebou\c{c}as, Jailson S. Alcaniz}
{Energy conditions constraints on a class of $f(R)$ gravity}

%
\catchline{}{}{}{}{}
%

\title{ENERGY CONDITIONS CONSTRAINTS ON A CLASS OF  $\mathbf{f(R)}\,$ GRAVITY}

\author{JANILO SANTOS}
\address{Universidade Federal do Rio G. do Norte\\ Departamento de F\'{\i}sica, C.P. 1641\\
59072-970 \ Natal -- RN, Brazil \\
janilo@dfte.ufrn.br}

\author{MARCELO J. REBOU\c{C}AS}
\address{Centro Brasileiro de Pesquisas F\'{\i}sicas\\ Rua Dr. Xavier Sigaud, 150\\
22290-180 \ Rio de Janeiro -- RJ, Brazil\\
reboucas@cbpf.br}

\author{JAILSON S. ALCANIZ}
\address{Observat\'orio Nacional\\ Departamento de Astronomia\\
20921-400\ Rio de Janeiro -- RJ, Brazil\\
  alcaniz@on.br}

\maketitle

\begin{history}
\received{Day Month Year}
\revised{Day Month Year}
\comby{Managing Editor}
\end{history}

\begin{abstract}
We present and discuss the bounds from the energy conditions
on a general $f(R)$ functional form in the framework of metric variational
approach. As a concrete application of the energy conditions to locally
homogeneous and isotropic $f(R)-$cosmology, the recent estimated values
of the deceleration and jerk parameters are used to examine the bounds
from the weak energy condition on the free parameter of the family of
$f(R)=\sqrt{R^2 - R_0^2}\;$ gravity theory.
\end{abstract}


\section{Introduction}
The observed late-time acceleration of the Universe poses a great
challenge to modern cosmology, which  may be the result of unknown
physical processes involving either modifications of gravitation
theory or the existence of new fields in high energy physics.
This latter route is most commonly used, however, following the former,
an attractive approach to this problem, known as $f(R)-$gravity,\cite{Kerner}
examines the possibility of modifying Einstein's general relativity (GR)
by adding terms proportional to powers of the Ricci scalar $R$ to the
Einstein-Hilbert Lagrangian (see also Refs.~\refcite{Francaviglia}
for recent reviews).
Although these theories provide an alternative way to explain the
observed cosmic acceleration without dark energy, the freedom
in the choice of different functional forms of $f(R)$ gives rise
to the problem of how to constrain on theoretical and/or
observational grounds, the many possible $f(R)-$gravity theories.
Theoretical limits have long been discussed in Refs.~\refcite{Amendola},
while only recently observational constraints from several cosmological
datasets have been explored for testing the viability of these
theories.\cite{Fabiocc}

Additional constraints to $f(R)$ theories may also arise by  imposing the
so-called energy conditions.\cite{Kung}  It is well known that these
conditions, initially formulated in GR context,\cite{Hawking} have been
used in different contexts to derive general results that hold for a
variety of physical situations.\cite{EC}
More recently, several authors have employed the GR classical energy conditions
to investigate cosmological issues such as the phantom fields\cite{Santos}
and the expansion history of the universe.\cite{Nilza} While they are well
founded in the context of GR, one has to be cautious when using the energy
conditions in a more general framework, such as the $f(R)-$gravity. In this
regard, in a recent work Santos \emph{et al.}\cite{SARC} have used
Raychaudhuri's equation along with the requirement that gravity is attractive,
to derive the energy conditions for a general $f(R)-$gravity in the
metric formulation.  They have shown that, although similar,
the energy conditions differ from their formulation in GR context.
In this work, we use estimated values of the deceleration and
jerk parameters to examine the bounds from these newly derived
$f(R)-$energy-conditions on the one-parameter family of a recently
proposed $f(R)-$gravity.\cite{Baghram-a,Baghram-b}

\section{Energy conditions in $\mathbf{f(R)}\,$ gravity}
The generalized action that defines an $f(R)-$gravity is given by
\begin{equation} \label{lagrangean}
 S=\int d^4x\sqrt{-g}\,f(R) + S_m
\end{equation}
where  $g=\det(g_{\mu\nu})$, $R$ is the Ricci scalar, $S_m$ is the
standard action for the matter fields, and hereafter we use units
such that $8\pi G=c=1$. Varying this action with respect to the metric
$g_{\mu\nu}$ we obtain the field equations
\begin{equation}  \label{field_eq}
f'R_{\mu\nu} - \frac{f}{2}g_{\mu\nu} - \left(\nabla_{\mu}\nabla_{\nu}-
g_{\mu\nu}\,\Box \,\right)f' = T_{\mu\nu}\,,
\end{equation}
where a prime denotes differentiation with respect to $R$ and
$\Box \equiv g^{\alpha \beta}\,\nabla_{\alpha}\nabla_{\beta}\,$.
Following and extending the GR approach, Santos \emph{et al.}\cite{SARC}
used the Eq.~(\ref{field_eq}) together with the Raychaudhury equation
and attractiveness of gravity to show that for the
Friedmann-Lema\^{\i}tre-Robertson-Walker (FLRW) geometry with scale factor
$a(t)$, the null energy condition (NEC) and the strong energy condition (SEC) in $f(R)-$gravity theories can be written in
the form
\begin{equation}  \label{NEC-f}
\mbox{\bf NEC} \quad \Longrightarrow \quad \rho + p  \geq 0\;, \quad
\mbox{and} \quad
\left( \ddot{R}-\dot{R}\,H \right)f'' + \dot{R}^2f''' \geq 0\,,
\end{equation}
\begin{equation}  \label{SEC-f}
\mbox{\bf SEC} \quad \Longrightarrow \quad \rho + 3p  - f + Rf'
+ 3\left( \ddot{R}+\dot{R}\,H\right)f'' + 3\dot{R}^2f''' \geq 0\,,
\end{equation}
where a dot denotes derivative with respect to time and $H=\dot{a}/a$
is the Hubble parameter. They also have  shown that, in addition to
Eq.~(\ref{NEC-f}),  the weak energy condition (WEC) requires
\begin{equation}  \label{WEC-f}
\mbox{\bf WEC} \quad \Longrightarrow \quad \rho + \frac{1}{2}(f - Rf')
- 3\dot{R}Hf'' \geq 0 \,,
\end{equation}
whereas the dominant energy conditions (DEC) fulfillment, besides the inequalities~(\ref{NEC-f})
and (\ref{WEC-f}), demands
\begin{equation}  \label{DEC-f}
\mbox{\bf DEC} \quad \Longrightarrow \quad \rho - p +  f - Rf'
- (\ddot{R}+ 5\dot{R}H)f'' - \dot{R}^2f''' \geq 0 \,.
\end{equation}
As one may easily check, for $f=R$, the well-known forms for the
NEC ($\rho + p  \geq 0$) and SEC ($\rho + 3p  \geq 0$) in the context
of GR can be recovered from Eqs.~(\ref{NEC-f}) and~(\ref{SEC-f}),
while Eqs.~(\ref{WEC-f}) and~(\ref{DEC-f}) give $\rho \geq 0$ and
$\rho - p\geq 0$, whose combination with Eqs~(\ref{NEC-f}) give,
respectively, the well-known forms of the WEC and DEC in GR
(see, e.g., Refs.~\refcite{Hawking} and \refcite{Nilza}).

\section{Constraining $\mathbf{f(R)=\sqrt{R^2 - R_0^2}}\,$ theory}
As shown by Santos \emph{et al.}\cite{SARC}, the energy-conditions inequalities~(\ref{NEC-f}), (\ref{SEC-f}),  (\ref{WEC-f})
and~(\ref{DEC-f}) can be used to place bounds on a given $f(R)$.
In the context of FLRW models the bounds can also be stated in
terms of the deceleration ($q$), jerk ($j$) and snap ($s$)
parameters. To this end, we note that  the Ricci scalar and
its derivatives can be expressed as
\label{eq:whole}
\begin{eqnarray}
R &=&-6H^2(1-q)\;, \\
\dot{R}&=&-6H^3(j-q-2)\;, \\
\ddot{R}&=&-6H^4(s+q^2+8q+6)\;, 
\end{eqnarray}
where
\begin{equation}
q=-\frac{1}{H^2}\frac{\ddot{a}}{a}\;, \qquad
j=\frac{1}{H^3}\frac{\dddot{a}}{a}\;,
\qquad s=\frac{1}{H^4}\frac{\ddddot{a}}{a}\;,
\end{equation}
and $H= \dot{a}/a$ is the Hubble parameter.
Thus, in terms of the present-day values (denoted by the subscript 0)
of the above parameters the {\bf NEC} [Eq.~(\ref{NEC-f})],
{\bf SEC} [Eq.~(\ref{SEC-f})], {\bf DEC} [Eq.~(\ref{DEC-f})] and
{\bf WEC} [Eq.~(\ref{WEC-f})]
can be, respectively, rewritten  as
\begin{equation} \label{nec0}
 \rho_0 + p_0 \geq 0 \quad \mbox{and} \quad
   -[s_0-j_0+(q_0+1)(q_0+8)]f''_0
+  6[H_0(j_0 -q_0 -2)]^2f'''_0 \geq 0,   \nonumber
\end{equation}
\begin{equation} \label{sec0}
\rho_0 +3 p_0 + f_0 - 6H^2_0(1-q_0)f'_0 -
6 H^4_0(s_0+j_0+q^2_0 + 7q_0+4)f''_0  + 3[6H^3_0(j_0 - q_0 -2)]^2f'''_0
\geq 0, \nonumber
\end{equation}
\begin{equation}  \label{dec0}
\rho_0 - p_0 + 6H^2_0(1-q_0)f'_0 - 6H^4_0[s_0
+ (q_0-1)(q_0+4) +5j_0]f''_0  - [6H^3_0(j_0-q_0-2)]^2f'''_0  \geq 0, \nonumber
\end{equation}
\begin{equation} \label{wec0}
 2 \rho_{0} + f_0 + 6 H^2_0(1-q_0)f'_0
+ 36 H^4_0(j_0-q_0-2)f''_0   \geq 0,
\end{equation}
which is in the appropriate form to confront with observations by using the
estimate values of the deceleration, jerk  and snap parameters.

Recently Baghram \emph{et al.}\cite{Baghram-a} and Movahed
\emph{et al.}\cite{Baghram-b} have proposed a modified $f(R)-$gravity by
choosing the geometric part of the Lagrangian (\ref{lagrangean}) as
\begin{equation} \label{f_R}
 f(R)=\sqrt{R^2 - R_0^2}\,,
\end{equation}
where $R_0$ is a parameter to be adjusted by observations.%
\footnote{Note that $R_0$ is not the present-day value of the
curvature scalar $R$.}
One features of this theory, is that it has an intrinsic minimum spatial
curvature, which provides a late time accelerating cosmic expansion.
Besides, its appealing mono-parameter form makes it more tractable
when testing for cosmological and solar system observations. Comparing
the observed perihelion precession of Mercury with the predictions for
this theory by  solving the equations~(\ref{field_eq}) for a spherically
symmetric Schwarzschild-type metric,  Baghram {\em et al.}\cite{Baghram-a}
have put an upper limit $R_0 < H^2_0$.

In order to examine how the energy conditions can be used to place
bounds on the free parameter $R_0$ of Eq.~(\ref{f_R}), we first note that,
apart from the WEC [Eq.~(\ref{wec0})], all above inequalities depend
on the current value of the snap parameter $s_0$. Therefore, since no
reliable measurement of this parameter has been reported hitherto,
we shall focus on the \textbf{WEC} requirement [given by Eq.~(\ref{wec0})] in
the confrontation of the energy condition bounds on the $f(R)-$theory
(\ref{f_R}) with observational data.
For a negligible value of the present-day density $\rho_0$, a straightforward
calculation shows that the \textbf{WEC}-fulfillment inequality leads to the
following constraint on the free parameter of the $f(R)-$theory (\ref{f_R}):
\begin{equation}  \label{R_0}
 R_0 \geq 6H^2_0\left( q_0^2 - 3q_0 -1 + j_0 \right)^{1/2}\;.
\end{equation}
Now, taking $q_0=-0.81\pm$ 0.14 and $j_0=2.16^{+0.81}_{-0.75}$, as given in
Ref.~\refcite{Rapetti}, we find that $R_0\geq 12.36 H^2_0$, which is
not consistent with the above-mentioned bound $R_0 < H^2_0$. In this way
the bound from the  perihelion precession of Mercury leads to a violation
of the \textbf{WEC}.%
\footnote{We note that the \textbf{WEC}-fulfilment bound on $R_0$ is not
very sensitive precision of the parameters $q_0$ and $j_0$, so future
more precise estimates of these parameters will not change considerably this
bound.}
%

\section{Final Remarks}

$f(R)-$gravity provides an alternative way to explain the current
cosmic acceleration with no need of invoking either the existence
of an extra spatial dimension or an exotic component of dark energy.
However, the arbitrariness in the choice of different functional
forms of $f(R)$  gives rise to the problem of how to constrain the
many possible $f(R)-$gravity theories on physical grounds. In this
work, following the Ref.\refcite{SARC}, we have showed how to put
constraints on general $f(R)-$gravity from the energy conditions
in the context of  locally homogeneous and isotropic
$f(R)-$cosmology.
In particular, for a negligible present-day density $\rho_0$, the
lower bound $R_0\geq 12.36 H^2_0$ arise from the fulfillment of
\textbf{WEC} for the $f(R)-$theory given by Eq.~(\ref{f_R}).
This \textbf{WEC}-fulfillment bound, however, is not consistent with the
bound on $R_0$ from the  perihelion precession of Mercury, making clear
the violation of the  \textbf{WEC} in this case.

\section*{Acknowledgments}
J. Santos acknowledges financial support from PRONEX (CNPq/FAPERN).
This work is supported by Conselho Nacional de Desenvolvimento
Cient\'{\i}fico e Tecnol\'{o}gico (CNPq) - Brasil, under grant No. 472436/2007-4.
M.J. Rebou\c{c}as and J.S. Alcaniz also thank CNPq for the grants under which
this work was carried out.


\end{document}